\documentclass{article}

\usepackage{arxiv}

\usepackage[utf8]{inputenc} % allow utf-8 input
\usepackage[T1]{fontenc}    % use 8-bit T1 fonts
\usepackage{hyperref}       % hyperlinks
\usepackage{url}            % simple URL typesetting
\usepackage{booktabs}       % professional-quality tables
\usepackage{amsfonts}       % blackboard math symbols
\usepackage{nicefrac}       % compact symbols for 1/2, etc.
\usepackage{microtype}      % microtypography
\usepackage{lipsum}		% Can be removed after putting your text content
\usepackage{graphicx}
\usepackage[square, numbers, sort&compress]{natbib}
\usepackage{doi}
\usepackage{amsmath}
\usepackage[ruled,vlined]{algorithm2e}
\usepackage{multirow}

\SetKw{Continue}{continue}
\SetKw{Or}{or}
\SetKw{And}{and}
\SetKw{Not}{not}

\title{Parallel Heuristic Exploration for Additive Complexity Reduction in Fast Matrix Multiplication}

\author{\href{https://orcid.org/0000-0001-8047-0114}{\includegraphics[scale=0.06]{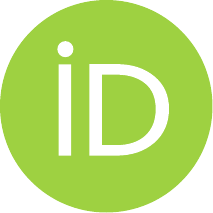}\hspace{1mm}Andrew I.~Perminov}\\
	Research Center for TAI\\
	Institute for System Programming\\
	Moscow \\
	\texttt{perminov@ispras.ru}
}

\renewcommand{\headeright}{}
\hypersetup{
pdftitle={Parallel Heuristic Exploration for Additive Complexity Reduction in Fast Matrix Multiplication},
pdfsubject={cs},
pdfauthor={Andrew I.~Perminov},
pdfkeywords={Fast matrix multiplication, Additive complexity reduction, Common subexpression elimination, Ternary coefficients set},
}

\begin{document}
\maketitle

\begin{abstract}
This paper presents a parallel random-search method for reducing additive complexity in fast matrix multiplication algorithms with ternary coefficients $\{-1,0,1\}$. The approach replaces expensive exact evaluation with fast heuristic scoring, including the new Greedy-Intersections strategy. The method runs many independent common subexpression elimination processes in parallel, exploring the search space through random pair substitutions and diverse selection strategies while sharing promising partial solutions. Tested on 149 ternary-coefficient schemes, the method achieves lower addition counts than the state-of-the-art Greedy-Potential on 102 schemes (including 57 new best-known results for optimal-rank schemes), matches it on 45, and is outperformed on only 2. For most schemes, it provides equal or better results while being significantly faster, making it practical for algorithm exploration. All software and results are open source.
\end{abstract}

% keywords can be removed
\keywords{Fast matrix multiplication \and Additive complexity reduction \and Common subexpression elimination \and Ternary coefficients set}

\section{Introduction}
After finding a fast matrix multiplication scheme with low rank, a second important problem appears: reducing the number of additions and subtractions needed to compute it. This additive complexity often determines real-world speed. Common subexpression elimination (CSE) is the standard method for this optimization, but it is a difficult combinatorial problem for the complex expression graphs of fast matrix multiplication algorithms.

This work focuses exclusively on schemes with coefficients in the ternary set $Z_T = \{-1,0,1\}$. Ternary schemes are particularly valuable for hardware implementation due to their simple coefficient representation, and they are also well-suited for testing addition reduction methods. The restriction to $Z_T$ allows for efficient data representations and specialized heuristics while maintaining relevance to practical fast matrix multiplication.

Recent work, such as the extensive study by Martensson et al~\citep{maartensson2025number}, has established benchmarks and introduced methods like ``Greedy-Potential'' to automate this process. However, the CSE problem for matrix multiplication remains difficult. Greedy strategies often get stuck in local optima, while methods that exactly evaluate each substitution are slow for large schemes.

This paper presents a parallel heuristic approach to additive complexity reduction for ternary-coefficient schemes. The method addresses the combinatorial challenge by running many independent CSE processes concurrently, each guided by different selection strategies. Two implementations are developed: a GPU-accelerated version that runs thousands of processes concurrently for maximum exploration, and a CPU version using OpenMP that runs tens to hundreds of processes for practical use. Both use the same core heuristic strategies.

The main contributions are:

\begin{itemize}
    \item A family of heuristics for CSE in fast matrix multiplication, with the new \emph{Greedy-Intersections} strategy as the main component. This heuristic quickly scores candidate substitutions using a stochastic function, balancing immediate gain and future potential without expensive full evaluations.
    
    \item A parallel optimization tool available in both GPU (CUDA) and CPU (OpenMP) versions. The tool runs many independent CSE processes concurrently. Processes use different strategies and share promising partial solutions, allowing wide exploration of the search space.
    
    \item Comprehensive evaluation on 149 known ternary-coefficient schemes. The method achieves lower addition counts than the \texttt{fmm\_add\_reduction}~\citep{fmm_add_reduction2025} baseline on 102 schemes, matches it on 45, and is outperformed on 2. 

    \item For 57 schemes with optimal known rank, the method achieves new reference addition counts, advancing the state-of-the-art for practical matrix multiplication implementations. All software and results are open source.
\end{itemize}

The results show that parallel heuristic search, combining randomness and diverse strategies, can find better reductions than purely greedy or local methods. This advances toward fully automated optimization of matrix multiplication code.

The paper is organized as follows: Section~\ref{sec:background} reviews related work. Section~\ref{sec:methodology} describes the methodology. Section~\ref{sec:experiments} presents experimental results. Section~\ref{sec:discussion} discusses limitations and implications. Section~\ref{sec:conclusion} concludes. Section~\ref{sec:availability} describes the availability of software and data.

\section{Background and Related Work}
\label{sec:background}
\subsection{Additive Complexity in Fast Matrix Multiplication}
The discovery of a matrix multiplication scheme with low rank is the primary goal for reducing multiplicative complexity. However, to achieve the best possible performance from a given scheme, a secondary optimization is essential: minimizing its additive complexity -- the number of required additions and subtractions.

For two schemes with the same rank, the one with fewer additions will generally execute faster, provided other factors like memory access patterns and cache utilization are comparable. In practical implementations, these architectural considerations also affect runtime, but they are beyond the scope of this paper, which focuses exclusively on minimizing the arithmetic operation count. Therefore, after finding a good low-rank scheme, applying common subexpression elimination (CSE) to minimize its additive cost becomes a critical optimization step. The naive addition count serves as a starting point, while the optimized count, achieved through CSE, represents the real arithmetic cost. For the irregular expression structures in fast matrix multiplication, CSE is a challenging combinatorial optimization problem where even small improvements are valuable, especially for schemes with an already optimal rank, as they yield new state-of-the-art algorithms.

\subsection{Prior Work on Additive Minimization}
The systematic study of additive minimization has a rich history, particularly for the foundational $2\times2\times2$ case. Strassen's original algorithm required 18 additions~\citep{strassen1969gaussian}. Winograd reduced this to 15~\citep{winograd1971multiplication}, which was long believed optimal until Karstadt and Schwartz demonstrated that a change of basis could further reduce the count to 12~\citep{karstadt2020matrix}. This breakthrough showed that the framework for optimality itself could be expanded, yielding significant practical speedups. The approach of finding new, sparser algebraic representations via basis change, as further developed by Beniamini et al.~\citep{beniamini2020sparsifying} and Holtz et al.~\citep{holtz2025alternative}, constitutes one major strategy for additive reduction.

The second major strategy is common subexpression elimination (CSE), which operates directly on a fixed scheme's computational graph. Tools like PLinOpt~\citep{dumas2024place} provide a general framework for optimizing linear, bilinear, and trilinear programs, including the generation of compact straight-line programs from matrix multiplication schemes. This work handles general integer and rational coefficients and can produce in-place accumulation programs. For the specific and highly structured problem of minimizing additions in ternary-coefficient schemes, more specialized tools have been developed. The state-of-the-art in automated CSE for this domain is the Greedy-Potential algorithm introduced by Martensson et al.~\citep{maartensson2025number}. This method, implemented in the \texttt{fmm\_add\_reduction} tool~\citep{fmm_add_reduction2025}, scores a candidate substitution by its ''potential'': the number of new substitution possibilities created after applying it. This calculation requires a full trial substitution and re-analysis of all two-element subexpressions, which becomes computationally expensive for large schemes.

These strategies represent distinct approaches to the additive complexity problem. Recent work by Stapleton~\citep{stapleton202560} demonstrates a different direction: instead of optimizing additions for an existing scheme, a neural network with ternary regularization discovers a new low-rank $3\times3\times3$ tensor decomposition from scratch. The resulting scheme is then post-processed with Greedy-Potential CSE, achieving a record of 60 additions for a rank-23 scheme. A similar principle -- searching for better algebraic forms before CSE optimization -- underlies the flip-graph~\citep{kauers2023flip} exploration in previous work, where ternary flips~\citep{perminov2025fast} were used to find schemes with minimal naive addition counts before applying CSE.

This paper focuses on improving the CSE strategy itself. While Greedy-Potential is effective, its need for expensive calculations restricts the search scope. The proposed approach instead uses parallel heuristic search. It replaces exact scoring with faster approximate heuristics, including a new Greedy-Intersections strategy. This avoids computational bottlenecks, enabling wider exploration of the search space.

\section{Methodology}
\label{sec:methodology}
\subsection{Matrix Multiplication Schemes}

A matrix multiplication scheme for multiplying matrices $A \in \mathbb{F}^{m \times n}$ and $B \in \mathbb{F}^{n \times p}$ over a field $\mathbb{F}$ with rank $r$ is defined by three coefficient tensors $U \in \mathbb{F}^{r \times m \times n}$, $V \in \mathbb{F}^{r \times n \times p}$, and $W \in \mathbb{F}^{r \times m \times p}$. It computes $r$ intermediate scalar products:

\begin{align*}
m_1 = (u^{(1)}_{11} a_{11} + \dots + u^{(1)}_{mn} a_{mn})\;\cdot\; & (v^{(1)}_{11} b_{11} + \dots + v^{(1)}_{np} b_{np})\\
\vdots\;  \\
m_r = (u^{(r)}_{11} a_{11} + \dots + u^{(r)}_{mn} a_{mn})\;\cdot\; & (v^{(r)}_{11} b_{11} + \dots + v^{(r)}_{np} b_{np}),
\end{align*}

and reconstructs the result matrix $C$, where $C = AB$, as:

\begin{align*}
c_{ij} = w^{(1)}_{ij}m_1 + \dots + w^{(r)}_{ij}m_r.
\end{align*}

The tensors $(U, V, W)$ must satisfy the Brent equations~\citep{brent1970algorithms}. A scheme is denoted by its format and rank as $(m, n, p: r)$. This work focuses on schemes where all coefficients in $U, V, W$ are restricted to the set $\{-1, 0, 1\}$, which are particularly efficient for hardware implementation.

\subsection{The Addition Minimization Problem}
The additive cost of a scheme originates from computing three independent sets of linear combinations:

\begin{itemize}
    \item \textbf{Input combinations for matrix A} ($E_U$): $r$ expressions, each combining $m \times n$ variables $a_{ij}$:
    \[
        e^{(l)}_U = u^{(l)}_{11} a_{11} + \cdots + u^{(l)}_{mn} a_{mn},\quad\text{for }l=1,\dots, r.
    \]

    \item \textbf{Input combinations for matrix B} ($E_V$): $r$ expressions, each combining $n \times p$ variables $b_{jk}$.
    \[
        e^{(l)}_V = v^{(l)}_{11} b_{11} + \cdots + v^{(l)}_{np} b_{np},\quad\text{for }l=1,\dots, r.
    \]

    \item \textbf{Output reconstruction for matrix C} ($E_W$): $m \cdot p$ expressions, each combining $r$ variables $m_l$ (the intermediate products):
    \[
        e^{(ij)}_W = w^{(1)}_{ij} m_1 + \cdots + w^{(r)}_{ij} m_r,\quad\text{for }i=1,\dots, m,j=1,\cdots,p.
    \]
\end{itemize}

The total additive complexity is the sum of the optimized operation counts for these three independent sets:
\[\text{Cost}(E_U) + \text{Cost}(E_V) + \text{Cost}(E_W).\]

The optimization of each set is a common, independent task.

\subsubsection{Common Subexpression Elimination Task}
The optimization of each set $E_X$ (where $X \in \{U, V, W\}$) is formulated as a common subexpression elimination problem.

Let $E$ be a set of $n_e$ linear expressions in $n_x$ input variables ${x_1, x_2, \dots, x_{n_x}}$:

\[
    E = \{e_1, e_2, \dots, e_{n_e}\},\quad\text{where } e_i = \sum_{j=1}^{n_x}{\alpha_{ij} \cdot x_j},\quad\alpha_{ij} \in \{-1, 0, 1\}. 
\]

The \textit{naive addition} count for computing $E$ directly is:

\[
    A_\text{naive}(E) = \sum_{i=1}^{n_e}\left({\sum_{j=1}^{n_x}\left|\alpha_{ij}\right|} - 1\right).
\]

The CSE process iteratively introduces fresh variables ${x_{n_x +1}, x_{n_x+2}, \dots, x_{n_x + n_f}}$. Each fresh variable $x_k$ is defined as a two-term linear combination of existing variables ($x_1, \dots, x_{k-1}$):

\[
    x_k = x_p \pm x_q\quad\text{for some } x_p, x_q \in \{x_1, x_2, \dots, x_{n_x}, x_{n_x+1}, \dots, x_{k-1}\}.
\]

Upon introducing $x_k$, the set of expressions is updated. All occurrences of the subexpression $(x_p \pm x_q)$ within the expressions of $E$ are replaced by the single variable $x_k$. This transforms the coefficient matrix: for each expression $e_i$, the coefficients for $x_p$ and $x_q$ are set to zero, and a new coefficient $\alpha_{ik} = \pm1$ for $x_k$ is introduced.

If the pattern $(x_p \pm x_q)$ occurs $c$ times across all expressions, the substitution saves $c - 1$ additions: one addition is spent to define $x_k$, and the all $c$ occurrences are replaced by a single variable access.

Let $E^{(final)}$ be the expression set after all substitutions, now defined over $n_x + n_f$ variables. Its naive cost is $A_{\text{naive}}(E^{(final)})$. The total optimized cost is the sum of the additions used to define all fresh variables plus this final naive cost:

\[
    Cost(E) = n_f + A_\text{naive}\left(E^{(final)}\right).
\]

The goal of the CSE algorithm is to find a sequence of pair substitutions that gives the smallest total cost $\text{Cost}(E)$.

\subsubsection{Illustrative Example}
Consider three expressions in variables ${x_1, x_2, x_3, x_4}$:

\begin{align*}
    e_1&=x_1+ x_2- x_3+ x_4, \\
    e_2&=x_1- x_2- x_4, \\
    e_3&=x_1- x_2- x_3+ x_4.
\end{align*}

The naive addition count is:
\[
    A_{\text{naive}} = (4-1)+(3-1)+(4-1)=8.
\]

Introduce fresh variables $x_5 = x_2 + x_4$ and $x_6 = x_1 - x_3$. After substitution, the system becomes:

\begin{align*}
    x_5&=x_2+ x_4, \\
    x_6&=x_1-x_3, \\
    e_1&=x_5+x_6, \\
    e_2&=x_1-x_5, \\
    e_3&=-x_2+x_4+x_6.
\end{align*}

The final expression set $E^{(final)}$ consists of ${e_1, e_2, e_3}$ over variables ${x_1, x_2, x_3, x_4, x_5, x_6}$.

Its naive cost is:
\[
    A_\text{naive}\left(E^{(final)}\right) = (2-1) + (2 - 1) + (3 - 1) = 4.
\]

Applying the cost formula with $n_f = 2$:

\[
    Cost(E) = n_f + A_\text{naive}\left(E^{(final)}\right) = 2 + 4 = 6.
\]

This represents a reduction from the original 8 to 6 additions, demonstrating a successful application of CSE. The example also shows the updated system, where $x_5$ and $x_6$ become new variables in the final expressions.

\subsection{Core algorithm}
The CSE algorithm optimizes each linear expression set $E$ (representing $E_U$, $E_V$, or $E_W$) independently through an iterative process. The procedure for a single set $E$ is described below.

\subsubsection{Algorithmic Overview}
The core optimization is an iterative loop that terminates when no beneficial substitution can be found. Algorithm~\ref{alg:cse} shows the complete procedure.

In the Analysis step, pairs are canonized: terms are ordered by index ($i < j$) and the sign is normalized so the coefficient of $x_i$ is $+1$. This ensures equivalent forms like $-x_i - x_j$ map to $(x_i + x_j)$. The Selection step uses one of the heuristic strategies described in Section~\ref{subsec:heuristic-strategies}. The loop repeats until no pair appears more than once.

\begin{algorithm}[H]
\caption{Common Subexpression Elimination (CSE) algorithm}
\label{alg:cse}
\DontPrintSemicolon
\KwIn{set of linear expressions $E = \{e_1, \dots, e_{n_e}\}$ in variables $\{x_1, \dots, x_{n_x}\}$}
\KwOut{optimized expressions $E^{(final)}$, fresh variables list $F$, total cost}
$F \leftarrow \emptyset$\;
$E_{current} \leftarrow E$\;
\BlankLine
\Repeat{no pair with frequency $> 1$ exists}{
    \textbf{Analysis:} find all canonical pairs $(x_i \pm x_j)$ in $E_{current}$ and count frequency $c$ for each pair\;
    \BlankLine
    \textbf{Selection:} choose pair $(x_i \pm x_j)$ with $c > 1$ using heuristic strategy\;
    \BlankLine
    \textbf{Substitution:}\;
    Create fresh variable $x_k$ with definition $x_k = x_i \pm x_j$\;
    $F \leftarrow F \cup \{x_k\}$\;
    \ForEach{$e \in E_{current}$}{
        Replace $(x_i \pm x_j)$ with $x_k$\;
        Replace $(-x_i \mp x_j)$ with $-x_k$\;
    }
}
\BlankLine
$E^{(final)} \leftarrow E_{current}$\;
$\text{Cost} \leftarrow |F| + A_{\text{naive}}(E^{(final)})$\;
\Return $E^{(final)}, F, \text{Cost}$\;
\end{algorithm}

\subsubsection{Algorithmic Scope and Rationale}
The algorithm only substitutes pairs of variables. This is enough because any longer common subexpression can be built step by step from pairs. For example, first replace $x_i + x_j$ with a new variable $x_{t_1}$. Then replace $x_{t_1} + x_k$ with $x_{t_2}$. This captures the longer expression $x_i + x_j + x_k$ in two steps.

The main challenge is choosing which pair to substitute next. Always picking the most frequent pair is simple but often leads to poor results. Better selection strategies are needed to explore the search space effectively. The next section describes several strategies, including the new Greedy-Intersections heuristic.

\subsection{Heuristic Selection Strategies}
\label{subsec:heuristic-strategies}

The pair selection strategy chooses which common subexpression to eliminate next. Different strategies explore the search space in different ways, often resulting in different final addition counts. This section describes the heuristic strategies used in the parallel search.

\subsubsection{Basic Strategies}

Four basic strategies provide different balances between exploitation and exploration:

\begin{itemize}
    \item \textbf{Greedy (g)}: always selects the pair with highest frequency $c$. This maximizes immediate gain but often gets stuck in local optima.
    
    \item \textbf{Greedy-Alternative (ga)}: finds all pairs with maximum frequency $c_{\text{max}}$, then selects one uniformly at random. This breaks ties randomly while still exploiting high-frequency pairs.
    
    \item \textbf{Weighted-Random (wr)}: selects a pair with probability proportional to its profit $c - 1$. More common pairs are more likely to be chosen, but any pair may be selected.
    
    \item \textbf{Greedy-Random (gr)}: with probability $p > 0.5$, selects a pair using the Greedy-Alternative strategy. Otherwise, selects a pair using the Weighted-Random strategy. This balances exploitation of high-frequency pairs with exploration of other options.
\end{itemize}

\subsubsection{The Greedy-Intersections Heuristic (gi)}
The Greedy-Potential method computes the value of a candidate pair by doing a trial substitution. It counts how many new substitution opportunities appear after the change. This evaluation is slow because it must re-analyze all expressions for each candidate pair.

The Greedy-Intersections heuristic estimates the value of a pair $s_p = (x_i \pm x_j)$ without doing the substitution. It uses only the current counted pair frequencies. The idea is based on how substitution affects other pairs.

When a pair $s_p$ is replaced by a fresh variable $x_k$, the effect on another pair $s_q$ depends on two factors: whether $s_p$ and $s_q$ share variables, and whether $s_q$ appears in expressions that contain $s_p$.

Pairs without intersection ($s_p \cap s_q = \emptyset$) are unaffected. They keep their current frequency. Pairs with intersection are affected only if they appear in the same expressions as $s_p$. If $s_q$ appears in expressions that do not contain $s_p$, it stays unchanged.

Exact tracking of expression overlap would require full expression analysis. Instead, the heuristic uses a stochastic approximation. For each intersecting pair $s_q$, it randomly decides whether $s_q$ will survive the substitution. The probability and weight of survival are determined empirically.

The score for pair $s_p$ is:

\[
H(s_p) = \underbrace{(c_{s_p} - 1)}_{\text{immediate gain}} +\ \alpha \cdot \sum_{s_q \neq s_p} I(s_p, s_q),
\]

where $\alpha$ is a scaling factor chosen randomly from $[0, 0.5]$ at the start of each reduction process, and $I(s_p, s_q)$ estimates the future value of $s_q$:

\[
I(s_p, s_q) = 
\begin{cases}
c_{s_q} - 1, & \text{if } s_p \cap s_q = \emptyset \\
0 \text{ or } \beta \cdot (c_{s_q} - 1) \text{ (each with probability 0.5)}, & \text{if } s_p \cap s_q \neq \emptyset
\end{cases}
\]

Here $\beta$ is a fixed value chosen randomly from $[0.5, 1.0]$ at the start of each reduction process. This stochastic approach accounts for the uncertainty in how intersecting pairs evolve after substitution.

The heuristic chooses the pair with maximum $H(s_p)$. Computing all scores requires checking pairwise intersections, which takes $O(m^2)$ time for $m$ pairs. This is faster than Greedy-Potential, which needs $O(m^2 \cdot n_e)$ time for $n_e$ expressions.

\subsubsection{Mixed Strategy (mix)}

The Mixed strategy selects between available strategies at each step according to predefined weights. The probability of selecting each strategy is proportional to its weight: Greedy-Intersections (weight 8), Greedy-Alternative (weight 4), Greedy-Random (weight 2), Weighted-Random (weight 1). This allows the optimization to adapt to different stages of the search process.

\subsection{Parallel Optimization Tool}

The CSE problem has many local optima. A single heuristic often finds a good but not optimal solution. To explore the search space more thoroughly, a parallel optimization tool is implemented in both GPU (CUDA) and CPU (OpenMP) versions.

\subsubsection{Parallel Reduction Process}

For a given matrix multiplication scheme, $N$ independent reduction processes are executed in parallel. In the GPU version, $N$ typically ranges from 2048 to 8192 processes, in the CPU version, from 64 to 256 processes. Each process reduces the three expression sets ($E_U$, $E_V$, $E_W$) independently using the iterative CSE algorithm.

The first process in the first iteration always uses the Greedy strategy as baseline. The remaining processes select their strategy randomly according to configurable weights. The default weights are: Greedy-Intersections (8), Greedy-Alternative (4), Greedy-Random (2), Weighted-Random (1), Mixed (0.1), and Greedy-Potential (0.01). This distribution favors the most promising heuristics while still exploring less common strategies. Users can adjust these weights for specific optimization tasks.

\subsubsection{Solution Sharing Mechanism}

From the second iteration onward, the tool identifies the best solution found so far (the substitution sequence with the lowest addition count). Then, a configurable percentage (default 40\%) of the reduction processes are partially reinitialized. Each reinitialized process copies the first $k$ substitutions from the best solution, where $k$ is chosen randomly between 1 and $3/4$ of the length of the best sequence. The process then continues optimizing from that partial solution.

This allows promising substitution patterns to propagate while maintaining search diversity.

\subsubsection{Dynamic Scheme Transformation via Random Flips}

As an additional exploration mode, the GPU tool can incorporate transformations from the ternary flip graph~\citep{perminov2025fast}. This mode was used during initial scheme discovery to find algebraic forms with potential for better additive reduction, though not in the final comparative experiments. The mode operates as follows:

\begin{enumerate}
    \item Given an input scheme $S$, create $M$ schemes $S_1, \ldots, S_M$ (typically $M = 32$).
    \item Scheme $S_1$ is an exact copy of the original scheme $S$.
    \item Each scheme $S_i$ for $i = 2,\ldots,M$ is created by applying a random number of ternary flip operations to $S$.
    \item These $M$ schemes are distributed to the $N$ parallel reduction processes.
    \item After the first optimization iteration completes, the process repeats: schemes $S_2,\ldots,S_M$ are regenerated by applying new random flip sequences to the original scheme $S$.
    \item The regenerated schemes are redistributed to reduction processes for the next iteration.
\end{enumerate}

This approach continuously explores fresh random transformations of the original scheme throughout the optimization. Each flip operation preserves mathematical correctness while creating different coefficient patterns in the $U$, $V$ and $W$ tensors. Different patterns may offer better opportunities for common subexpression elimination.

The mode explores many alternative schemes with the same rank, searching for forms that are easier to optimize. The best result is a scheme with lower additive complexity while keeping the same rank.

\subsection{Implementation Details}
\label{subsec:implementation}

The implementation uses a compact representation for linear expressions. An expression $e = \alpha_1 x_{i_1} + \dots + \alpha_t x_{i_t}$ with coefficients $\alpha_j \in \{-1, 1\}$ is stored as a set of signed integers. For a variable $x_i$:
\begin{itemize}
    \item if coefficient $+1$: store $+i$,
    \item if coefficient $-1$: store $-i$.
\end{itemize}
Thus expression $x_1 - x_4 + x_5 - x_9$ is stored as $\{1, -4, 5, -9\}$.

During CSE, substituting a fresh variable $x_k$ for the pair $(x_i \pm x_j)$ replaces all occurrences of $(i, \pm j)$ with $+k$ and all occurrences of $(-i, \mp j)$ with $-k$ in the signed integer representation. This requires removing the corresponding signed entries from the set and inserting $+k$ or $-k$. Set operations (insertion, deletion, membership test) are implemented in $O(1)$ average time using hash tables. Therefore, replacing a pair across all expressions costs $O(n_e)$ operations where $n_e$ is the number of expressions.

This representation is memory-efficient for ternary coefficients and enables fast pair identification and substitution, supporting the parallel execution of thousands of independent CSE processes.

\section{Experimental results}
\label{sec:experiments}
\subsection{Experimental Setup}
\label{subsec:setup}

The proposed parallel optimization tool is evaluated on 149 fast matrix multiplication schemes with ternary coefficients. The schemes include both previously known algorithms and new discoveries obtained through ternary flip graph exploration~\citep{perminov2025fast}. Starting schemes were selected to have minimal naive addition counts as favorable starting points for additive reduction.

The baseline for comparison is the \texttt{fmm\_add\_reduction} tool~\citep{fmm_add_reduction2025}. Three configurations of this baseline are tested:
\begin{itemize}
    \item \textbf{Greedy-Vanilla}: simple greedy strategy (GV);
    \item \textbf{Greedy-Potential (5 steps)}: default greedy-potential mode with 5 values of the $\alpha$ parameter linearly spaced from 0 to 0.5 (GP-5);
    \item \textbf{Greedy-Potential (40 steps)}: greedy-potential with 40 values of $\alpha$ linearly spaced from 0 to 0.5, representing a more exhaustive search (GP-40).
\end{itemize}

Two versions of the proposed tool are tested: a GPU version and a CPU version. The GPU version adapts the number of parallel processes to memory constraints:
\begin{itemize}
    \item \textbf{Small schemes} (rank $r < 100$): 16,384 parallel processes;
    \item \textbf{Medium schemes} ($100 \leq r < 200$): 8,192 parallel processes;
    \item \textbf{Large schemes} ($r \geq 200$): 2,048 parallel processes.
\end{itemize}

The CPU version runs on an Intel(R) Core(TM) i7-8700 CPU @ 3.20GHz and uses fewer processes:
\begin{itemize}
    \item \textbf{Small schemes}: 256–512 parallel processes;
    \item \textbf{Medium and large schemes}: 32–64 parallel processes.
\end{itemize}

Each optimization runs until convergence: the best addition count remains unchanged for 10 consecutive iterations. The total number of iterations ranges from 10-15 for small schemes to 50-100 for larger schemes. GPU experiments were performed on an NVIDIA GTX 1650 laptop GPU and an RTX 2050 desktop GPU, both with 6GB memory.

\subsection{Additive Complexity Reduction Results}

Across 149 ternary-coefficient schemes, the results demonstrate two key outcomes: the establishment of new state-of-the-art addition counts for optimal-rank schemes, and a general ability to find better reductions than exhaustive search methods across a wide set of problems.

\subsubsection{New State-of-the-Art Addition Counts}

A primary practical achievement is the improvement of 57 matrix multiplication schemes that already have the best-known (optimal) rank for their dimensions. For these algorithms, which represent a multiplicative complexity bound, proposed method finds a more efficient evaluation scheme, setting a new reference for their total arithmetic cost. Table~\ref{tab:sota_new} lists of these new SOTA results.

\begin{table}[ht!]
	\caption{Schemes with optimal rank achieving new SOTA addition counts}
    \label{tab:sota_new}
    \small
	\centering
	\begin{tabular}{ccccccc|ccccccc}
        \toprule
        Format & Rank & \multirow{2}{*}{Naive} & \multirow{2}{*}{GV} & GP & GP & \multirow{2}{*}{Prop.} & Format & Rank & \multirow{2}{*}{Naive} & \multirow{2}{*}{GV} & GP & GP & \multirow{2}{*}{Prop.} \\
        $m \times n \times p$ & $r$ & & & 5 & 40 & & $m \times n \times p$ & $r$ & & & 5 & 40 \\
        \midrule
        $2\times3\times10$ & 50 & 198  & 135  & 135  & 135  & \textbf{134} & $4\times6\times7$  & 123 & 1586 & 535  & 533  & 518  & \textbf{517} \\
$2\times3\times13$ & 65 & 256  & 170  & 170  & 170  & \textbf{169} & $4\times6\times8$  & 140 & 1248 & 576  & 558  & 558  & \textbf{551} \\
$2\times4\times7$  & 45 & 308  & 181  & 178  & 177  & \textbf{174} & $4\times6\times10$ & 175 & 1878 & 759  & 738  & 723  & \textbf{715} \\
$2\times4\times8$  & 51 & 354  & 197  & 191  & 190  & \textbf{188} & $4\times7\times8$  & 164 & 1505 & 752  & 729  & 698  & \textbf{690} \\
$2\times4\times9$  & 59 & 309  & 210  & 208  & 208  & \textbf{207} & $4\times8\times8$  & 182 & 1884 & 861  & 812  & 800  & \textbf{795} \\
$2\times4\times11$ & 71 & 430  & 275  & 271  & 268  & \textbf{265} & $5\times5\times5$  & 93 & 843  & 392  & 391  & 387  & \textbf{383} \\
$2\times4\times12$ & 77 & 484  & 282  & 277  & 276  & \textbf{274} & $5\times5\times6$  & 110 & 1215 & 491  & 479  & 466  & \textbf{460} \\
$2\times4\times14$ & 90 & 616  & 326  & 320  & 318  & \textbf{313} & $5\times5\times7$  & 127 & 1607 & 561  & 553  & 539  & \textbf{531} \\
$2\times4\times15$ & 96 & 662  & 366  & 358  & 357  & \textbf{351} & $5\times5\times8$  & 144 & 1924 & 645  & 638  & 631  & \textbf{620} \\
$2\times4\times16$ & 102 & 708  & 352  & 344  & 342  & \textbf{338} & $5\times5\times9$  & 167 & 1814 & 706  & 699  & 690  & \textbf{682} \\
$2\times5\times5$  & 40 & 283  & 156  & 154  & 154  & \textbf{153} & $5\times5\times10$ & 184 & 2116 & 796  & 787  & 781  & \textbf{774} \\
$2\times5\times6$  & 47 & 332  & 189  & 186  & 184  & \textbf{181} & $5\times5\times11$ & 202 & 2272 & 879  & 872  & 859  & \textbf{850} \\
$2\times5\times9$  & 72 & 465  & 275  & 270  & 270  & \textbf{266} & $5\times5\times12$ & 220 & 2444 & 848  & 835  & 811  & \textbf{799} \\
$2\times5\times11$ & 87 & 540  & 331  & 326  & 326  & \textbf{323} & $5\times6\times6$  & 130 & 1716 & 589  & 587  & 573  & \textbf{562} \\
$2\times5\times12$ & 94 & 664  & 336  & 330  & 328  & \textbf{322} & $5\times6\times7$  & 150 & 2039 & 696  & 691  & 671  & \textbf{664} \\
$2\times6\times9$  & 86 & 548  & 302  & 296  & 295  & \textbf{293} & $5\times6\times8$  & 170 & 2312 & 747  & 740  & 728  & \textbf{720} \\
$2\times6\times10$ & 94 & 668  & 334  & 327  & 327  & \textbf{325} & $5\times6\times9$  & 197 & 2376 & 884  & 875  & 850  & \textbf{840} \\
$3\times5\times5$  & 58 & 357  & 230  & 224  & 223  & \textbf{221} & $5\times6\times10$ & 217 & 2772 & 956  & 954  & 940  & \textbf{925} \\
$4\times4\times5$  & 61 & 452  & 247  & 237  & 237  & \textbf{233} & $5\times7\times7$  & 176 & 2605 & 832  & 820  & 799  & \textbf{788} \\
$4\times4\times6$  & 73 & 540  & 288  & 282  & 281  & \textbf{280} & $5\times8\times8$  & 230 & 2747 & 1019 & 1010 & 980  & \textbf{974} \\
$4\times4\times7$  & 85 & 631  & 326  & 323  & 320  & \textbf{319} & $6\times6\times6$  & 153 & 2182 & 704  & 685  & 671  & \textbf{655} \\
$4\times4\times8$  & 96 & 973  & 396  & 391  & 387  & \textbf{377} & $6\times6\times7$  & 183 & 2502 & 810  & 790  & 777  & \textbf{769} \\
$4\times5\times6$  & 90 & 1023 & 401  & 393  & 386  & \textbf{380} & $6\times6\times8$  & 203 & 1994 & 896  & 880  & 868  & \textbf{836} \\
$4\times5\times7$  & 104 & 931  & 423  & 413  & 405  & \textbf{400} & $6\times6\times9$  & 225 & 2440 & 1029 & 951  & 936  & \textbf{923} \\
$4\times5\times8$  & 118 & 1521 & 543  & 538  & 522  & \textbf{513} & $6\times7\times7$  & 215 & 2004 & 965  & 900  & 886  & \textbf{880} \\
$4\times5\times10$ & 151 & 1207 & 590  & 580  & 571  & \textbf{566} & $6\times7\times8$  & 239 & 2263 & 1112 & 1050 & 1042 & \textbf{1027} \\
$4\times5\times11$ & 165 & 1801 & 718  & 702  & 689  & \textbf{681} & $6\times7\times9$  & 268 & 3062 & 1184 & 1166 & 1152 & \textbf{1146} \\
$4\times5\times12$ & 179 & 1977 & 778  & 772  & 761  & \textbf{750} & $6\times8\times8$  & 266 & 2780 & 1244 & 1214 & 1181 & \textbf{1161} \\
$4\times6\times6$  & 105 & 894  & 467  & 435  & 435  & \textbf{430} &  \\
    \bottomrule
	\end{tabular}
\end{table}

\subsubsection{General Improvements Across All Schemes}
Beyond establishing new SOTA records, the parallel heuristic search demonstrates high effectiveness in the additive minimization task. It finds a lower addition count than the exhaustive GP-40 method for an additional 45 schemes that do not have optimal rank. Table~\ref{tab:general_improved} shows a representative sample of these improvements, confirming the method's ability to escape local optima encountered by other strategies.

\begin{table}[ht!]
	\caption{Representative general improvements (non-optimal rank schemes)}
    \label{tab:general_improved}
    \small
	\centering
	\begin{tabular}{ccccccc|ccccccc}
        \toprule
        Format & Rank & \multirow{2}{*}{Naive} & \multirow{2}{*}{GV} & GP & GP & \multirow{2}{*}{Prop.} & Format & Rank & \multirow{2}{*}{Naive} & \multirow{2}{*}{GV} & GP & GP & \multirow{2}{*}{Prop.} \\
        $m \times n \times p$ & $r$ & & & 5 & 40 & & $m \times n \times p$ & $r$ & & & 5 & 40 \\
        \midrule
$2\times4\times13$ & 84  & 595  & 323  &     316      &      316      & \textbf{313}  & $3\times6\times10$ & 140 & 1152 & 494  &     484      &      477      & \textbf{476}  \\
$2\times5\times7$  & 57  & 340  & 197  &     194      &      192      & \textbf{189}  & $3\times7\times7$  & 115 & 789  & 441  &     434      &      434      & \textbf{432}  \\
$2\times5\times8$  & 65  & 376  & 222  &     218      &      218      & \textbf{214}  & $3\times7\times9$  & 147 & 1002 & 592  &     583      &      580      & \textbf{579}  \\
$2\times6\times7$  & 68  & 396  & 233  &     230      &      230      & \textbf{226}  & $4\times4\times4$  & 49  & 474  & 167  &     164      &      163      & \textbf{159}  \\
$2\times6\times8$  & 77  & 456  & 272  &     269      &      269      & \textbf{266}  & $4\times4\times9$  & 110 & 925  & 445  &     424      &      422      & \textbf{419}  \\
$2\times7\times8$  & 90  & 648  & 350  &     333      &      331      & \textbf{329}  & $4\times4\times10$ & 122 & 879  & 493  &     482      &      482      & \textbf{480}  \\
$2\times7\times9$  & 102 & 678  & 398  &     384      &      382      & \textbf{379}  & $4\times4\times11$ & 134 & 1150 & 534  &     528      &      524      & \textbf{519}  \\
$3\times3\times6$  & 43  & 284  & 168  &     164      &      164      & \textbf{160}  & $4\times4\times12$ & 145 & 1495 & 619  &     606      &      596      & \textbf{590}  \\
$3\times3\times9$  & 65  & 347  & 218  &     216      &      216      & \textbf{215}  & $4\times4\times13$ & 157 & 1474 & 650  &     638      &      635      & \textbf{628}  \\
$3\times3\times12$ & 86  & 582  & 303  &     297      &      297      & \textbf{289}  & $4\times4\times14$ & 169 & 1596 & 711  &     702      &      694      & \textbf{691}  \\
$3\times3\times14$ & 101 & 664  & 373  &     365      &      363      & \textbf{361}  & $4\times4\times15$ & 181 & 1679 & 741  &     738      &      735      & \textbf{729}  \\
$3\times4\times6$  & 56  & 359  & 211  &     210      &      210      & \textbf{209}  & $4\times4\times16$ & 192 & 2059 & 730  &     728      &      724      & \textbf{718}  \\
$3\times4\times10$ & 93  & 582  & 351  &     346      &      346      & \textbf{345}  & $4\times5\times9$  & 137 & 1217 & 564  &     555      &      547      & \textbf{542}  \\
$3\times4\times15$ & 138 & 958  & 544  &     537      &      535      & \textbf{533}  & $4\times6\times9$  & 160 & 1472 & 720  &     651      &      647      & \textbf{646}  \\
$3\times5\times6$  & 70  & 561  & 270  &     267      &      267      & \textbf{265}  & $4\times7\times7$  & 145 & 1381 & 680  &     646      &      632      & \textbf{629}  \\
$3\times5\times7$  & 83  & 494  & 315  &     308      &      305      & \textbf{303}  & $4\times7\times9$  & 187 & 2059 & 806  &     800      &      788      & \textbf{785}  \\
$3\times5\times9$  & 105 & 646  & 390  &     384      &      384      & \textbf{383}  & $5\times7\times8$  & 206 & 1880 & 948  &     892      &      882      & \textbf{879}  \\
$3\times5\times11$ & 128 & 967  & 497  &     489      &      488      & \textbf{484}  & $5\times7\times9$  & 231 & 2554 & 993  &     981      &      971      & \textbf{963}  \\
$3\times5\times12$ & 140 & 1152 & 491  &     483      &      483      & \textbf{482}  & $7\times7\times7$  & 250 & 2417 & 1119 &     1077     &     1070      & \textbf{1067} \\
$3\times6\times6$  & 85  & 998  & 367  &     364      &      359      & \textbf{352}  & $7\times7\times8$  & 279 & 2926 & 1369 &     1260     &     1248      & \textbf{1237} \\
$3\times6\times7$  & 100 & 704  & 364  &     352      &      351      & \textbf{350}  & $7\times7\times9$  & 316 & 3452 & 1460 &     1404     &     1385      & \textbf{1383} \\
$3\times6\times8$  & 113 & 1240 & 467  &     459      &      446      & \textbf{445}  & $8\times8\times8$  & 343 & 4434 & 1748 &     1709     &     1668      & \textbf{1661} \\
$3\times6\times9$  & 127 & 1095 & 519  &     502      &      493      & \textbf{489}  &  \\
		\bottomrule
	\end{tabular}
\end{table}

In total, the proposed method achieves a strict improvement over GP-40 for 102 of the 149 tested schemes (68.4\%). The complete list of all 102 improved schemes is available in the included open-source repository.

\subsubsection{Equal Performance}
The proposed method achieves the same addition count as GP-40 for 45 schemes. These results are summarized in two categories, based on the rank of the scheme.

Table~\ref{tab:equal_optimal} shows cases where the scheme's rank matches the current best known in the literature. For many of these  schemes, no previously published addition count was found to serve as a direct benchmark, apart from those cataloged by Martensson et al.~\citep{maartensson2025number}. Therefore, when the proposed method matches the result from the state-of-the-art GP-40 tool on these schemes, it confirms that both approaches converge to the same high-quality solution. This shared result likely represents the current best-known (and possibly optimal) addition count for that particular matrix multiplication algorithm.

\begin{table}[ht!]
	\caption{Equal addition counts (proposed method matches GP-40) for schemes with best-known rank}
    \label{tab:equal_optimal}
	\centering
	\begin{tabular}{cccccccc}
        \toprule
        Format & Rank & Best & \multirow{2}{*}{Naive} & \multirow{2}{*}{GV} & \multirow{2}{*}{GP-5} & \multirow{2}{*}{GP-40} & \multirow{2}{*}{Proposed} \\
        $m \times n \times p$ & $r$ & known & & & & & \\
        \midrule
$2\times3\times3$  & 15  &  48~\citep{maartensson2025number}  &  58  &  46  &      46      &      46       &      46       \\
$2\times3\times4$  & 20  &  99~\citep{maartensson2025number}  &  82  &  58  &      58      &      58       &      58       \\
$2\times3\times5$  & 25  & 103~\citep{maartensson2025number}  & 110  &  72  & \textbf{71}  &  \textbf{71}  &  \textbf{71}  \\
$2\times3\times6$  & 30  &  -   & 116  &  81  &      81      &      81       &      81       \\
$2\times3\times7$  & 35  &  -   & 140  & 100  & \textbf{99}  &  \textbf{99}  &  \textbf{99}  \\
$2\times3\times8$  & 40  &  -   & 164  & 104  &     104      &      104      &      104      \\
$2\times3\times9$  & 45  &  -   & 174  & 116  &     116      &      116      &      116      \\
$2\times3\times11$ & 55  &  -   & 222  & 146  & \textbf{145} & \textbf{145}  & \textbf{145}  \\
$2\times3\times12$ & 60  &  -   & 232  & 151  &     151      &      151      &      151      \\
$2\times3\times14$ & 70  &  -   & 280  & 181  & \textbf{180} & \textbf{180}  & \textbf{180}  \\
$2\times3\times15$ & 75  &  -   & 307  & 192  &     192      &      192      &      192      \\
$2\times3\times16$ & 80  &  -   & 328  & 196  &     196      &      196      &      196      \\
$2\times4\times4$  & 26  & 173~\citep{maartensson2025number}  & 130  &  93  & \textbf{92}  &  \textbf{92}  &  \textbf{92}  \\
$2\times4\times6$  & 39  &  -   & 202  & 138  & \textbf{136} & \textbf{136}  & \textbf{136}  \\
$2\times8\times8$  & 100 &  -   & 608  & 427  & \textbf{424} & \textbf{424}  & \textbf{424}  \\
$3\times3\times3$  & 23  &  60~\citep{stapleton202560}  &  97  &  60  &      60      &      60       &      60       \\
$3\times3\times4$  & 29  & 105~\citep{maartensson2025number}  & 134  &  93  & \textbf{92}  &  \textbf{92}  &  \textbf{92}  \\
$3\times3\times5$  & 36  & 176~\citep{maartensson2025number}  & 193  & 125  & \textbf{123} & \textbf{123}  & \textbf{123}  \\
$3\times4\times4$  & 38  & 198~\citep{maartensson2025number}  & 194  & 136  & \textbf{133} & \textbf{133}  & \textbf{133}  \\
$3\times4\times5$  & 47  & 276~\citep{maartensson2025number}  & 277  & 170  & \textbf{161} & \textbf{161}  & \textbf{161}  \\
$3\times5\times10$ & 115 &  -   & 730  & 391  & \textbf{380} & \textbf{380}  & \textbf{380}  \\
$4\times5\times5$  & 76  & 451~\citep{maartensson2025number}  & 530  & 315  &     300      & \textbf{299}  & \textbf{299}  \\
		\bottomrule
	\end{tabular}
\end{table}

Table~\ref{tab:equal_other} presents schemes where the rank is above the best known. For these, the addition count is optimized for a non-optimal multiplication count.

\begin{table}[ht!]
	\caption{Equal addition counts (proposed method matches GP-40) for schemes with non-optimal rank}
    \label{tab:equal_other}
	\small
	\centering
	\begin{tabular}{ccccccc|ccccccc}
        \toprule
        Format & Rank & \multirow{2}{*}{Naive} & \multirow{2}{*}{GV} & GP & GP & \multirow{2}{*}{Prop.} & Format & Rank & \multirow{2}{*}{Naive} & \multirow{2}{*}{GV} & GP & GP & \multirow{2}{*}{Prop.} \\
        $m \times n \times p$ & $r$ & & & 5 & 40 & & $m \times n \times p$ & $r$ & & & 5 & 40 \\
        \midrule
$2\times4\times5$  & 33  & 184  & 114  & \textbf{112} & \textbf{112}  & \textbf{112}  & $3\times4\times7$  & 64  & 446  & 252  &     251      & \textbf{249}  & \textbf{249}  \\
$2\times4\times10$ & 65  & 340  & 223  & \textbf{222} & \textbf{222}  & \textbf{222}  & $3\times4\times8$  & 74  & 461  & 272  & \textbf{267} & \textbf{267}  & \textbf{267}  \\
$2\times5\times10$ & 80  & 416  & 275  & \textbf{273} & \textbf{273}  & \textbf{273}  & $3\times4\times9$  & 84  & 535  & 324  &     317      & \textbf{316}  & \textbf{316}  \\
$2\times6\times6$  & 57  & 326  & 231  & \textbf{228} & \textbf{228}  & \textbf{228}  & $3\times4\times11$ & 102 & 641  & 387  &     384      & \textbf{381}  & \textbf{381}  \\
$2\times7\times7$  & 77  & 452  & 323  & \textbf{320} & \textbf{320}  & \textbf{320}  & $3\times4\times12$ & 111 & 721  & 424  &     412      & \textbf{411}  & \textbf{411}  \\
$3\times3\times7$  & 51  & 279  & 155  & \textbf{154} & \textbf{154}  & \textbf{154}  & $3\times4\times13$ & 121 & 804  & 471  &     456      & \textbf{455}  & \textbf{455}  \\
$3\times3\times8$  & 58  & 275  & 170  & \textbf{169} & \textbf{169}  & \textbf{169}  & $3\times4\times14$ & 128 & 896  & 466  &     460      & \textbf{458}  & \textbf{458}  \\
$3\times3\times10$ & 72  & 386  & 217  & \textbf{214} & \textbf{214}  & \textbf{214}  & $3\times4\times16$ & 148 & 1043 & 532  & \textbf{518} & \textbf{518}  & \textbf{518}  \\
$3\times3\times11$ & 79  & 493  & 296  &     287      & \textbf{286}  & \textbf{286}  & $3\times5\times8$  & 94  & 584  & 295  &     295      &      295      &      295      \\
$3\times3\times13$ & 94  & 593  & 341  &     335      & \textbf{334}  & \textbf{334}  & $3\times7\times8$  & 128 & 930  & 462  & \textbf{450} & \textbf{450}  & \textbf{450}  \\
$3\times3\times15$ & 108 & 579  & 309  & \textbf{305} & \textbf{305}  & \textbf{305}  & $3\times8\times8$  & 148 & 1020 & 519  & \textbf{511} & \textbf{511}  & \textbf{511}  \\
$3\times3\times16$ & 115 & 862  & 424  &     410      & \textbf{407}  & \textbf{407}  &  \\
		\bottomrule
	\end{tabular}
\end{table}

In most of these 45 cases, GP-5 and GP-40 produce identical addition counts, suggesting these values may represent optimal or near-optimal solutions for the given algebraic form. The consistent 35--45\% reduction from naive counts across these schemes further demonstrates the effectiveness of CSE optimization.

\subsubsection{Inferior Performance}
For only 2 of the 149 tested schemes does the exhaustive GP-40 method find a lower addition count than the proposed parallel heuristic. These cases are shown in Table~\ref{tab:worse}. Notably, for both schemes, the proposed method still achieves a lower count than the practical default GP-5 configuration.

\begin{table}[ht!]
	\caption{Schemes where GP-40 performs better than proposed method}
    \label{tab:worse}
	\centering
	\begin{tabular}{ccccccc}
        \toprule
        Format & Rank & Naive & GV & GP-5 & GP-40 & Proposed \\
        \midrule
        $6\times6\times10$ & 252 & 3540 & 1291 & 1249 & \textbf{1167} & 1210 \\
        $7\times8\times8$  & 310 & 3604 & 1670 & 1498 & \textbf{1471} & 1494 \\
		\bottomrule
	\end{tabular}
\end{table}

The performance gap is most pronounced for very large schemes. The largest difference is 43 additions for $(6,6,10:252)$. Achieving this result with GP-40 requires evaluating 40 different $\alpha$ parameter values -- a process that can take tens of hours for such large schemes. In contrast, the parallel heuristic CPU search completes in 3–4 hours.

These results highlight the persistent combinatorial difficulty of additive complexity reduction. While exhaustive search can find slightly better solutions for the very largest schemes, it is much slower. The proposed method offers a better trade-off: it finds near-optimal solutions much faster for almost all schemes. It achieves near-optimal results significantly faster for the vast majority of schemes and still outperforms the practical default configuration (GP-5) in most challenging cases.

\subsubsection{Component-Wise Analysis}

An analysis of $U$, $V$ and $W$ components separately reveals that different optimization methods can find better reductions in different parts of the same scheme. Table~\ref{tab:component-wise} shows schemes where the proposed method and \texttt{fmm\_add\_reduction} achieve different addition counts for individual components. For these schemes, taking the minimum value from each component across both methods gives a total addition count lower than what either method achieves alone.

\begin{table}[ht!]
	\caption{Component-wise addition counts for schemes improvable by approaches result combination}
    \label{tab:component-wise}
	\centering
    \small
	\begin{tabular}{cc|ccc|ccc|ccc|ccc}
        \toprule
        Format & Rank & \multicolumn{3}{|c}{Additions of $U$} & \multicolumn{3}{|c}{Additions of $V$} & \multicolumn{3}{|c}{Additions of $W$} & \multicolumn{3}{|c}{Total additions} \\
        $m \times n \times p$ & $r$ & Prop. & GP-40 & Min & Prop. & GP-40 & Min & Prop. & GP-40 & Min & Prop. & GP-40 & Min \\
        \midrule
$4\times7\times8$  & \textbf{164} &  \textbf{158}  &      160       & 158  &  \textbf{227}  &      234       & 227  &      305       &  \textbf{304} & 304  & 690  & 698  & \textbf{689}  \\
$4\times8\times8$  & \textbf{182} &  \textbf{200}  &      205       & 200  &  \textbf{256}  &      258       & 256  &      339       &  \textbf{337} & 337  & 795  & 800  & \textbf{793}  \\
$5\times7\times8$  & 206 &      195       &  \textbf{194}  & 194  &      290       &      290       & 290  &  \textbf{394}  &      398      & 394  & 879  & 882  & 878  \\
$6\times6\times10$ & 252 &      320       &  \textbf{289}  & 289  &      370       &  \textbf{356}  & 356  &  \textbf{520}  &      522      & 520  & 1210 & 1167 & 1165 \\
$7\times7\times7$  & 250 &      288       &      288       & 288  &      290       &  \textbf{289}  & 289  &  \textbf{489}  &      493      & 489  & 1067 & 1070 & 1066 \\
$7\times8\times8$  & 310 &  \textbf{384}  &      387       & 384  &      454       &  \textbf{444}  & 444  &      656       &  \textbf{640} & 640  & 1494 & 1471 & 1468 \\
$8\times8\times8$  & 343 &  \textbf{462}  &      465       & 462  &  \textbf{456}  &      465       & 456  &      743       &  \textbf{738} & 738  & 1661 & 1668 & 1656 \\
		\bottomrule
	\end{tabular}
\end{table}

The columns "Min" show the minimum addition count achieved for each component across both methods. This reveals that:
\begin{itemize}
    \item The proposed method often finds better reductions for $U$ and $V$ components.
    \item \texttt{fmm\_add\_reduction} sometimes outperforms proposed method in optimizing the $W$ components.
    \item Neither method consistently dominates across all three components.
\end{itemize}

This analysis suggests potential for a \emph{hybrid approach}: combining the best-optimized $U$, $V$ and $W$ components from different optimization runs could achieve even lower total addition counts than any single method. Notably, for the schemes $(4,7,8:164)$ and $(4,8,8:182)$ -- which have optimal ranks -- this component-wise minimum creates new SOTA totals of 689 and 793 additions, respectively. These improve upon the totals found by either method alone for these schemes.

\subsubsection{Summary Statistics}

Across the 149 ternary-coefficient schemes tested, the proposed parallel method shows clear improvements over the current best method:
\begin{itemize}
    \item \textbf{102 schemes (68.4\%)}: proposed method achieves lower addition counts (including 57 new SOTA for optimal-rank schemes).
    \item \textbf{45 schemes (30.2\%)}: proposed method matches GP-40 results.
    \item \textbf{2 schemes}: GP-40 achieves better addition counts.
\end{itemize}

For the majority of cases, the parallel heuristic search provides equal or better results than the exhaustive GP-40 method while being significantly faster. On moderate-sized schemes, the proposed approach typically completes in minutes, while GP-40 can take hours.

For the very largest schemes (rank $>250$), the pattern changes. The proposed method may require several hours and can yield slightly higher addition counts than GP-40. This highlights the combinatorial difficulty of additive reduction: heuristic exploration works well when the search space is tractable but faces greater challenges as dimensionality grows.

Despite this limitation on the most extreme cases, the proposed method consistently outperforms the practical default configuration GP-5, achieving lower addition counts on 114 schemes (76.5\%). This performance, combined with its ability to establish new SOTA records, makes it a valuable tool for rapid exploration and optimization of matrix multiplication schemes.

\section{Discussion and Limitations}
\label{sec:discussion}

The parallel heuristic approach shows that massive exploration with diverse strategies, including randomness and occasional use of expensive methods like Greedy-Potential, can overcome local optima in common subexpression elimination. The success of the Greedy-Intersections heuristic, which avoids expensive trial substitutions, demonstrates that fast approximate scoring can effectively guide the search while maintaining computational efficiency.

\subsection{Limitations}

Several limitations of the current approach deserve attention:

\begin{itemize}
    \item \textbf{Combinatorial complexity}: as scheme size grows, the search space expands exponentially. Although the parallel tool explores many paths, it cannot guarantee optimality. Results show diminishing returns on large schemes, where exhaustive methods like GP-40 retain an advantage.

    \item \textbf{Heuristic dependence}: the quality of results relies on the set of heuristics used. While the current heuristics work well for smaller schemes, larger schemes with more complex expression graphs might benefit from different heuristic designs or adaptive strategies that learn during optimization.
    
    \item \textbf{Coefficient set and scope}: the method is developed and evaluated only for schemes with ternary coefficients in $\{-1,0,1\}$. Its effectiveness on schemes with other coefficient sets (such as integer $\mathbb{Z}$ or rational $\mathbb{Q}$) remains untested. Furthermore, this work focuses exclusively on minimizing the arithmetic operation count. Practical implementation performance also depends on factors such as memory access patterns and cache utilization, which are beyond the scope of this current optimization.
    
    \item \textbf{Randomness}: the stochastic nature means results are not deterministic. Multiple runs can produce different addition counts, though variance is generally low for most schemes.
\end{itemize}

\subsection{Practical Implications}

The parallel tool is fast, making it useful for developing matrix multiplication schemes. Researchers can quickly check the addition count of different schemes, allowing them to explore more options. The tool and results are open source, so others can use and improve them.

\subsection{Future Work}

Several directions for improvement emerge:
\begin{itemize}
    \item \textbf{Adaptive heuristics}: strategies that learn from the optimization process could improve performance on large schemes.
    \item \textbf{Hierarchical search}: combining the parallel approach with systematic search in promising regions might bridge the gap to exhaustive methods.
    \item \textbf{Integration with basis optimization}: joint optimization of basis and CSE could yield further additive reductions.
    \item \textbf{Extension to non-ternary coefficients}: generalizing the approach to other coefficient sets would make it useful for more schemes.

    \item \textbf{Practical implementation and analysis}: future work includes evaluating the practical performance of the optimized schemes. This involves analyzing their memory access patterns, conducting numerical stability experiments, and measuring actual runtime on target hardware to validate the real-world benefit of the reduced addition counts.
\end{itemize}

\section{Conclusion}
\label{sec:conclusion}

This paper introduced a parallel heuristic approach for reducing the additive complexity of fast matrix multiplication algorithms with ternary coefficients ($\{-1,0,1\}$). To overcome the combinatorial nature of common subexpression elimination, the method executes many independent optimization processes, guided by a diverse set of strategies including the novel \emph{Greedy-Intersections} heuristic.

Evaluated on 149 known ternary-coefficient schemes, the approach demonstrates that massive parallel exploration is highly effective. It achieves lower addition counts than the exhaustive state-of-the-art method (GP-40) on 102 schemes (68.4\%), matches it on 45 (30.2\%), and is outperformed on only 2. Notably, this includes establishing 57 new state-of-the-art addition counts for schemes with optimal rank, directly improving the total arithmetic cost of the best-known algorithms for these matrix dimensions.

The analysis revealed that different optimization methods often excel on different parts ($U$, $V$ and $W$) of a scheme, pointing to a clear opportunity for future hybrid optimizers. While exhaustive search retains an advantage for the very largest schemes, the parallel heuristic provides a superior balance of speed and solution quality for the vast majority of cases, making it a practical tool for algorithm exploration and development.

All software, results, and optimized schemes are released as open source, providing a foundation for further research and a valuable resource for the computational algebra community.

\section{Availability}
\label{sec:availability}

The complete implementation of the parallel optimization tools, supporting utilities, and all optimized matrix multiplication schemes are publicly available under open-source licenses:

\begin{itemize}
    \item \textbf{GPU parallel tool}: \url{https://github.com/dronperminov/FlipGraphGPU}. Source code for the GPU-accelerated additive reduction tool, including the Greedy-Intersections heuristic and parallel optimization engine.

    \item \textbf{CPU parallel tool}: \url{https://github.com/dronperminov/ternary_addition_reducer}. OpenMP-based version of the tool, suitable for CPU environments.
    
    \item \textbf{Research database}: \url{https://github.com/dronperminov/FastMatrixMultiplication}. Complete collection of all experimental results and optimized schemes, including:
    \begin{itemize}
        \item Results for all 149 ternary-coefficient schemes tested in this work;
        \item Comparison data against Greedy-Potential baselines (GP-5 and GP-40);
        \item Schemes with improved addition counts over previous records.
    \end{itemize}
\end{itemize}

The repositories include documentation, usage examples, and scripts to reproduce the experimental results. The tools can be used as standalone optimizers or integrated into larger scheme exploration workflows.

\bibliographystyle{unsrtnat}
\bibliography{references}  %%% Uncomment this line and comment out the ``thebibliography'' section below to use the external .bib file (using bibtex) .





% \end{thebibliography}

\end{document}